\documentstyle[aps,psfig,prl,multicol]{revtex}

\begin{document}
%\preprint{\Large \it Second draft}
\vspace{1.0cm}

\title{Nucleon Charge and Magnetization Densities}
\author{James J. Kelly}
\address{ Department of Physics, University of Maryland, 
          College Park, MD 20742 }
\date{Revised: April 22, 2002}
\maketitle

\begin{abstract}
We use a relativistic prescription to extract charge and magnetization
densities from data for the scattering of high-energy electrons by nucleons.
A Fourier-Bessel analysis is used to minimize the model dependence of the
fitted densities.
We find that the neutron and proton magnetization densities are very similar, 
but the proton charge density is significantly softer.
A useful measurement of the neutron charge density is obtained, although
the relative uncertainty in the interior will remain substantially larger than
for the other densities until precise new data at higher $Q^2$ become 
available.
\end{abstract}

\begin{multicols}{2}[]
\narrowtext
\vfill
\eject

The electromagnetic structure of nucleons can be investigated using the
electric and magnetic Sachs form factors, $G_E$ and $G_M$, measured
by elastic scattering of high-energy electrons.
Recent experiments with intense high-polarization electron beams have 
improved the quality of the data for nucleon elastic form factors.
Perhaps the most dramatic observation \cite{MKJones00,Gayou02}
is that the ratio between the electric and magnetic form factors for the 
proton decreases sharply for $1 < Q^2 < 6$ (GeV/$c$)$^2$.
It was suggested that these results demonstrate that the proton charge is
distributed over a larger volume than its magnetization, 
but radial densities were not obtained.  
At low momentum transfer, the form factors are related to Fourier 
transforms of the charge and magnetization densities, 
but this interpretation is more complicated at high momentum transfer.   
Nevertheless, although models of nucleon structure can often calculate
the form factors directly, it remains desirable to relate the form 
factors to spatial densities because our intuition tends to be based more
firmly in space than momentum transfer.
In this paper we use a Fourier-Bessel analysis, 
together with a relativistic relationship between form factors and densities,
to parametrize the nucleon electromagnetic form factors in terms of
charge and magnetization densities.

The nucleon electromagnetic vertex function takes the form
\begin{equation}
\Gamma^\mu = F_1(Q^2)\gamma^\mu + 
\kappa F_2(Q^2) \frac{i \sigma^{\mu\nu}q_\nu}{2m}
\end{equation}
where  $F_1$ and $F_2$ are known as Dirac and Pauli form factors,
$\kappa$ is the anomalous part of the magnetic moment, 
and $\gamma^\mu$ and $\sigma^{\mu\nu}$ are the usual Dirac matrices 
({\it e.g.}, \cite{BjorkenDrella}).
This current operator appears simplest in the Breit frame where the 
nucleon approaches with initial momentum $-\vec{q}_{B}/2$, receives 
three-momentum transfer $\vec{q}_B$, and leaves with final momentum 
$\vec{q}_B/2$ such that the energy transfer vanishes.
In the Breit frame for a particular value of $Q^2$, the current  
separates into electric and magnetic contributions \cite{Sachs62}
\begin{equation}
\label{eq:BreitCurrent}
\bar{u}(p^\prime,s^\prime) \Gamma^\mu u(p,s) = 
\chi^\dagger_{s^\prime}
\left( G_E + \frac{i\vec{\sigma}\times\vec{q}_B}{2m} G_M 
\right) \chi_s
\end{equation}
where $\chi_s$ is a two-component Pauli spinor and where the
Sachs form factors are given by
\begin{mathletters}
\begin{eqnarray}
G_E &=& F_1 - \tau \kappa F_2 \\
G_M &=& F_1 + \kappa F_2
\end{eqnarray}
\end{mathletters}
with $\tau=(Q/2m)^2$.
Naively it would appear that one could obtain charge and magnetization
densities as Fourier transforms of the electric and magnetic form factors,
but each momentum transfer $Q$ specifies a different Breit frame.
Early experiments with modest $Q^2$ suggested that 
\begin{displaymath}
G_{Ep} \approx \frac{G_{Mp}}{\mu_p} \approx \frac{G_{Mn}}{\mu_n}
\approx G_D
\end{displaymath}
where $\mu$ is the magnetic moment in nuclear magnetons and
$G_D(Q^2) = (1 + Q^2/\Lambda^2)^{-2}$ with 
$\Lambda^2 = 0.71$ (GeV/$c$)$^2$ is known as the dipole form factor.
Similarly, data for $G_{En}$ at low $Q^2$ can be described by the 
Galster parametrization \cite{Galster71}
\begin{equation}
\label{eq:Galster}
G_{En}(Q^2) \approx -\mu_n G_D(Q^2) \frac{A \tau}{1+B\tau}
\end{equation}
where $A$ and $B$ are constants.

Let $\rho_{ch}(r)$ and $\rho_{m}(r)$ represent spherical intrinsic charge
and magnetization densities normalized according to
\begin{displaymath}
\int dr \; r^2 \rho_{ch}(r) = Z, \hspace{0.25cm}
\int dr \; r^2 \rho_{m}(r) = 1
\end{displaymath}
%\begin{mathletters}
%\begin{eqnarray}
%\int dr \; r^2 \rho_{ch}(r) &=& Z \\
%\int dr \; r^2 \rho_{m}(r) &=& 1
%\end{eqnarray}
%\end{mathletters}
where $Z=0,1$ is the nucleon charge.
An {\it intrinsic form factor} can then be defined by the
Fourier-Bessel transform
\begin{equation}
\tilde{\rho}(k) = \int dr \; r^2 j_0(kr) \rho(r)
\end{equation}
where $k$ is the intrinsic spatial frequency (or wave number).
If one knew how to obtain $\tilde{\rho}(k)$ from data for the
appropriate Sachs form factor, the intrinsic density could be
obtained simply by inverting the Fourier transform, such that
\begin{equation}
\label{eq:rhor}
\rho(r) = \frac{2}{\pi} \int_0^\infty dk \; k^2 j_0(kr) \tilde{\rho}(k)
\end{equation}
The naive or nonrelativistic inversion from form factor to density
replaces $Q$ by $k$ and $\tilde{\rho}(k)$ by $G(Q^2)$,
but leads to unsatisfactory densities for dipole or Galster form
factors --- the direct Fourier transform of a dipole form factor is 
an exponential density with a cusp at the origin.
The corresponding density for the Galster form factor is more
complicated but also features an unphysical cusp at the origin. 

Although there is no rigorous model-independent relationship between
the Sachs form factor and the corresponding static density in the
rest frame, a variety of models offer prescriptions of the form
\begin{mathletters}
\label{eq:rhok}
\begin{eqnarray}
\tilde{\rho}_{ch}(k) &=& G_E(Q^2) (1+\tau)^{\lambda_E} \\
\mu \tilde{\rho}_{m}(k) &=& G_M(Q^2) (1+\tau)^{\lambda_M} 
\end{eqnarray}
\end{mathletters}
where $\lambda$ is a model-dependent constant and
where the intrinsic spatial frequency $k$ is related to the invariant
momentum transfer by
\begin{equation}
\label{eq:k}
k^2 = \frac{Q^2}{1+\tau}
\end{equation}
The most important relativistic effect is the Lorentz contraction of  
spatial distributions in the Breit frame and the corresponding increase
of spatial frequency represented by the 
factor of $(1+\tau)$ in Eq. (\ref{eq:k}).
A measurement with Breit-frame momentum transfer $q_B = Q$ probes a 
reduced spatial frequency $k$ in the rest frame.
The Sachs form factor for a large invariant momentum transfer $Q^2$ is 
determined by a much smaller spatial frequency $k^2=Q^2/(1+\tau)$ and 
thus declines much less rapidly with respect to $Q^2$ than the Fourier 
transform of the density declines with respect to $k^2$.
Licht and Pagnamenta \cite{Licht70b} demonstrated that by accounting
for Lorentz contraction a good fit to data for $G_{Ep}$ can obtained using 
a Gaussian density typical of quark models.
Thus, this procedure provides physically reasonable densities
free of cusps at the origin.
However, according to Eq. (\ref{eq:k}), the maximum spatial frequency
that is accessible with spacelike $Q^2$ is $k_m=2m$.
The limitation  $k<k_m$ can be interpreted as a consequence
of relativistic position fluctuations, known as {\it zitterbewegung},
which obscure details finer that the nucleon Compton wavelength.

The first attempt to relate elastic form factors to ground-state densities 
was made by Licht and Pagnamenta \cite{Licht70b} using a quark cluster model 
in the impulse approximation and a kinematic boost.
They proposed a relativistic inversion method using Eq. (\ref{eq:rhok})
with $\lambda_E=\lambda_M=1$.
However, these choices for $\lambda$ do not ensure that the Sachs form
factors scale with $Q^{-4}$, as expected from perturbative QCD, unless
restrictions are placed upon $\tilde{\rho}(k_m)$.
Using a more symmetric treatment of the cluster model, 
Mitra and Kumari \cite{Mitra77} found that $\lambda_E=\lambda_M=2$ 
automatically satisfy the perturbative QCD scaling relations at very 
large $Q^2$ without constraining $\tilde{\rho}(k_m)$.
More recently, Ji \cite{Ji91} obtained similar relationships with 
$\lambda_E=0$ and $\lambda_M=1$ using a relativistic Skyrmion 
model based upon a Lorentz invariant Lagrangian density for which the
classical soliton solution can be evaluated in any frame.
Quantum fluctuations were then evaluated after the boost.
Furthermore, the recent soliton calculation by Holzwarth \cite{Holzwarth96}
uses this prescription and accurately predicted the decline in 
$G_{Ep}/G_{Mp}$.
In this paper we employ the values $\lambda_E=0$ and $\lambda_M=1$
suggested by the soliton model and in a subsequent paper \cite{Kelly02a}
will investigate in detail the {\it discrete ambiguity} due to the 
choice of $\lambda$.  

To extract radial densities from the nucleon form factor data we 
employ techniques originally developed for fitting radial distributions 
to data for scattering of electrons or protons from nuclei
\cite{Dreher74,Friar73,Kelly88a}.
Simple models with a small number of parameters do not offer
sufficient flexibility to provide a realistic estimate of the uncertainty
in a radial density.
Rather, we employ linear expansions in complete sets of basis functions 
that are capable of describing any plausible radial distribution without 
strong {\it a priori} constraints upon its shape.
Such expansions permit one to estimate the uncertainties in the fitted
density due to both the statistical quality of the data and the 
inevitable limitation of experimental data to a frequency range,  
$k \leq k_{max}$.
The uncertainty due to limitation of $k$ is known as 
{\it incompleteness error}.
More detailed discussion of the method may be found in Refs. 
\cite{Dreher74,Friar73,Kelly88a},
but the basic idea is to supplement the experimental data by pseudodata
of the form $\tilde{\rho}(k_i) = 0 \pm \delta \tilde{\rho}(k_i)$
whose uncertainties are based upon a reasonable model of the
asymptotic behavior of the form factor for $k_i > k_{max}$
where $k_{max}$ is the spatial frequency corresponding to the 
maximum measured $Q^2$. 
On quite general grounds one expects the asymptotic form factor for a 
confined system to decrease more rapidly than $k^{-4}$
\cite{Friar73}.
Therefore, we assume that
\begin{equation}
\label{eq:env}
\delta\tilde{\rho}(k) = \tilde{\rho}(k_{max}) 
\left( \frac{k_{max}}{k} \right)^4
\end{equation}
Notice that we must apply this procedure to $\tilde{\rho}(k)$ instead
of the Sachs form factors because, according to Eq. (\ref{eq:k}), 
the accessible spatial frequencies are limited to $k < 2m$ for
spacelike $Q^2$ while convergence of Eq. (\ref{eq:rhor}) requires a
constraint upon $\tilde{\rho}(k)$ for inaccessible range $k>2m$.

The Fourier-Bessel expansion (FBE) takes the form
\begin{equation}
\rho(r) = \sum_n a_n j_0(k_n r) \Theta(R-r)
\end{equation}
where $\Theta$ is the unit step function, $R$ is the expansion radius,
$k_n=n\pi/R$ are the roots of the Bessel function, and $a_n$ are the
coefficients to be fitted to data.
One advantage of the FBE is that the contribution of each term to the form
factor is concentrated around its $k_n$ so that a coefficient $a_n$ is
largely determined by data with $k \sim k_n$.
The larger the expansion radius $R$, the smaller the spacing between successive
$k_n$ and the greater the sensitivity one has to variations in the form factor.
One should choose $R$ to be several times the root-mean-square radius
but not so large that an excessive number of terms is needed to span the 
experimental range of momentum transfer.
Terms with $k_n > k_{max}$ provide an estimate of the incompleteness error.
We chose $R = 4.0$ fm, but the results are insensitive to its exact value.
Small but undesirable oscillations in fitted densities at large radius
were suppressed using a {\it tail bias} based upon the method
discussed in Ref. \cite{Kelly91b}.
We employed a tail function of the form $t(r) \propto e^{-\Lambda r}$,
based upon the successful dipole parametrization for low $Q^2$, 
and included in the $\chi^2$ fit a penalty for strong deviations from
the tail function for $r>2.0$ fm.
The constraint on the neutron charge was also enforced using a 
penalty function.
The tail bias improves the convergence of moments of the density 
but has practically no effect upon a fitted density in the region where 
it is large. 
The error band for a fitted density is computed from the covariance matrix
for the $\chi^2$ fit and includes the incompleteness error, 
but does not include discrete ambiguities due to the choices for $\lambda$ 
in Eq. (\ref{eq:rhok}).

We selected the best available data in each range of $Q^2$, with an emphasis
upon recent data using recoil or target polarization wherever available.
$G_{Mp}$ data were taken from the compilation of H\"ohler \cite{Hohler76}
for $Q^2 < 0.15$ (GeV/$c$)$^2$ and for larger $Q^2$ from the analysis
of Brash {\it et al}.\ \cite{Brash01}
using the recent recoil polarization data for $G_{Ep}/G_{Mp}$ from  
Refs.\ \cite{MKJones00,Gayou02}.
Cross section data from Refs.\ \cite{Price71,Simon80} were used 
for $Q^2 < 1$ (GeV/$c$)$^2$ but cross section data for $G_{Ep}$ were
excluded for larger $Q^2$.
Similarly, the data for $G_{En}$ were limited to recent polarization data
\cite{Eden94b,Passchier99,Rohe99,Herberg99,Golak01,Zhu01},
with corrections for nuclear structure and final-state interactions
whenever available,
plus the analysis of $t_{20}$ and $T_{20}$
by Schiavilla and Sick \cite{Schiavilla01}. 
We also included the neutron charge radius obtained by 
Kopecky {\it et al}.\ \cite{Kopecky97}
from transmission of thermal neutron through liquid $^{208}$Pb and
$^{209}$Bi.
Finally, for $G_{Mn}$ we selected polarization data from \cite{Xu00} 
and cross section data from 
\cite{Rock82,Lung93,Markowitz93,Anklin94,Bruins95,Anklin98,Kubon02}.

\begin{figure}[ht]
\centerline{ \strut\psfig{file=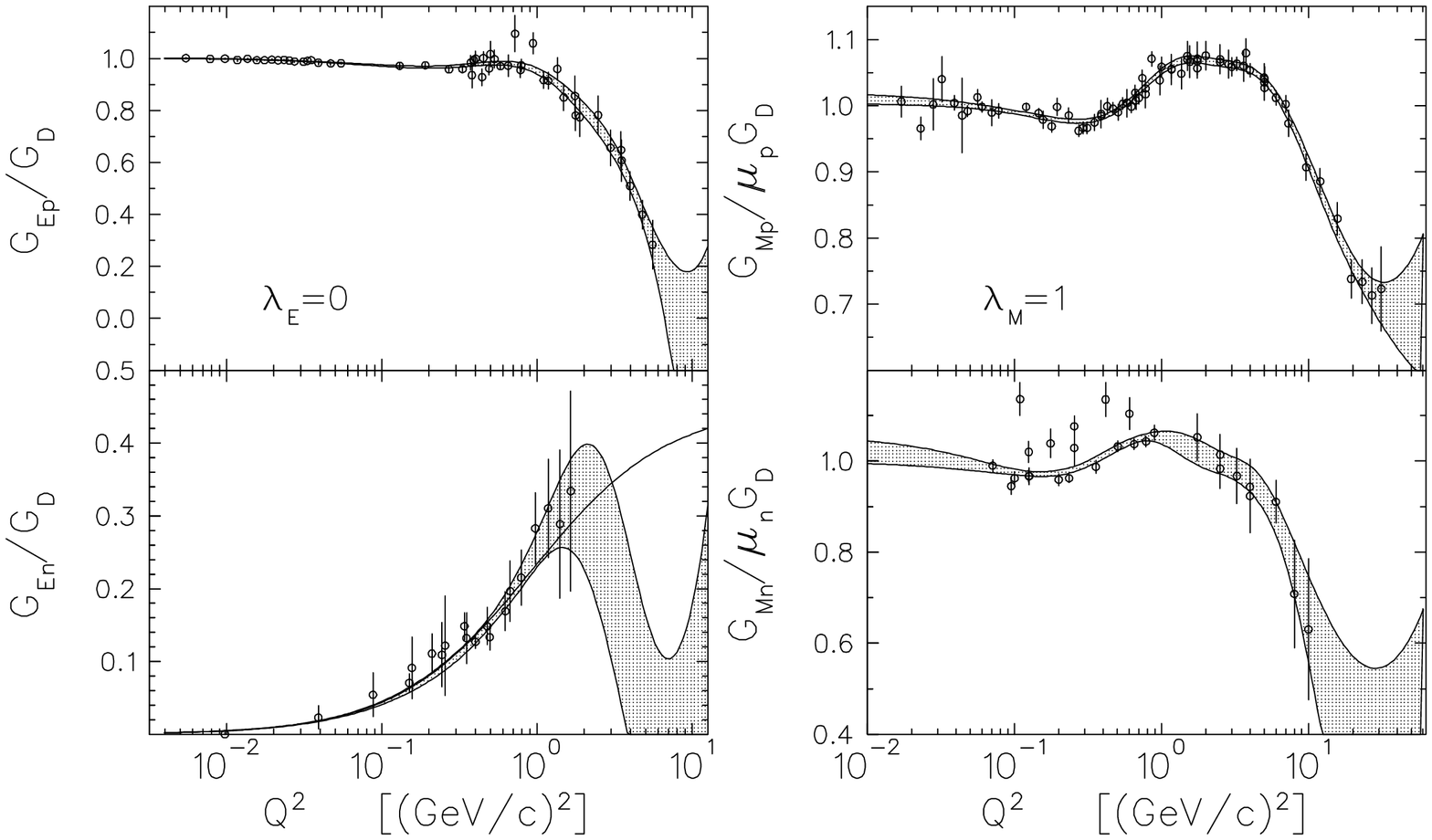,width=5.0in,angle=90} }
\caption{The bands show Fourier-Bessel fits to selected data for
nucleon electromagnetic form factors.
For $G_{En}$ the solid line shows the Galster model.}
\label{fig:ff}
\end{figure}

Fits to the form factor data are shown in Fig.\ \ref{fig:ff} as bands 
that represent the uncertainties in the fitted form factors. 
The widths of these bands are comparable to the experimental precision
where data are available, but expand for larger $Q^2$ where the 
uncertainties are based upon Eq. (\ref{eq:env}).
Notice, however, that in this figure the form factors were divided by $G_D$, 
which decreases rapidly with $Q^2$.
Therefore, the impact of uncertainties at large $Q^2$ upon the fitted 
densities remains modest because, with the exception of $G_{En}$, 
the form factors themselves and their absolute uncertainties become rather 
small at their largest $Q^2$.
Although the low-$Q^2$ data for $G_{Mn}$ have improved in recent years,
significant systematic discrepancies remain.
Recent data from Refs.\ \cite{Xu00,Anklin94,Anklin98,Kubon02}
with small statistical uncertainties suggest a small dip near 0.2 and a peak 
near 1 (GeV/$c$)$^2$.
For $G_{En}$ we plot the Galster model \cite{Galster71} for comparison.
The simple two-parameter fit Galster {\it et al.} made to the rather poor
data available at that time did not permit a realistic estimate of the 
uncertainty in the form factor or fitted density and the apparent agreement 
with more modern data must be judged as remarkable but fortuitous.

\begin{figure}[htbp]
\centerline{ \strut\psfig{file=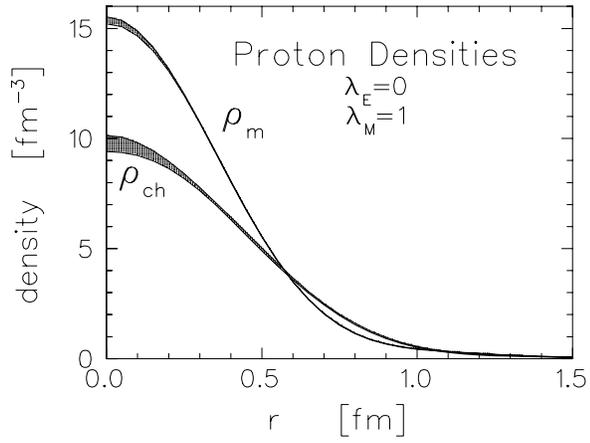,width=3.0in,angle=90} }
\caption{Comparison between fitted charge ($\rho_{ch}$) and 
magnetization ($\rho_m$) densities for the proton.  
Both densities are normalized to $\int dr \; r^2 \rho(r) = 1$.}
\label{fig:proton}
\end{figure}

\begin{figure}[hbtp]
\centerline{ \strut\psfig{file=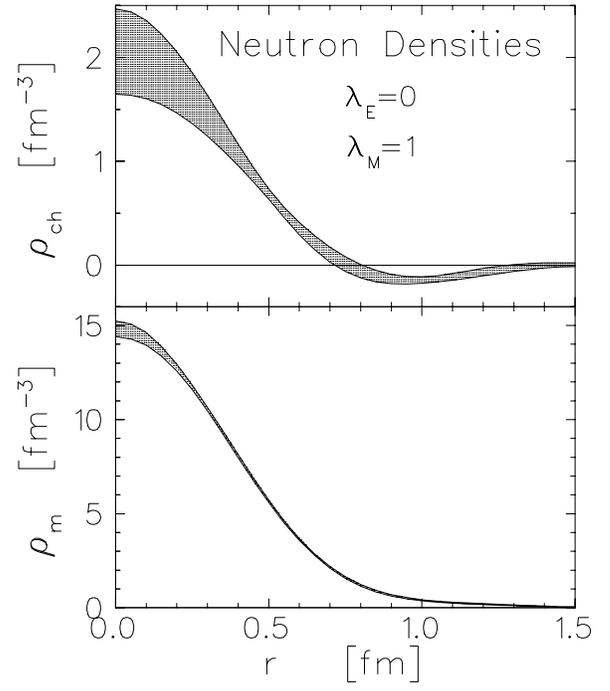,width=3.0in} }
\caption{Charge ($\rho_{ch}$) and magnetization ($\rho_m$) densities 
for the neutron.}
\label{fig:neutron}
\end{figure}

Proton charge and magnetization densities are compared in 
Fig.\ \ref{fig:proton}.
Both densities are measured very precisely.
The new recoil-polarization data for $G_{Ep}$ decrease more rapidly
than either the dipole form factor or the magnetic form factor
for $Q^2 > 1$ (GeV/$c$)$^2$; consequently, the charge density is
significantly softer than the magnetization density of the proton.
Neutron densities are shown in Fig. \ref{fig:neutron}.
We find that the magnetization density for the neutron is very similar to 
that for the proton,
although the interior precision is not as good because the range of $Q^2$ 
is smaller and the experimental uncertainties larger.
Limitations in the range and quality of the $G_{En}$ data presently
available result in a substantially wider error band for the neutron
charge density.
Data at higher $Q^2$ are needed to improve the interior precision,
but a useful measurement of the interior charge density is obtained 
nonetheless.
The positive interior density is balanced by a negative surface lobe.
Note that polarization measurements are sensitive to the sign of the density, 
but that cross section measurements are not.

In summary, we have developed a parametrization of nucleon electromagnetic
form factors based upon radial densities expanded in complete bases.
Although we cannot claim that the fitted densities are unique, this method 
does provide an intuitively appealing parametrization of the data.
It is necessary to account for Lorentz contraction at high momentum
transfer in order to obtain reasonable interior densities without a
cusp at the origin.
Using recent recoil polarization data for the proton, 
we find that the charge density is significantly broader than the 
magnetization density.
We also find that the neutron and proton magnetization densities
are similar.
Although the available data are not yet as precise, the neutron charge
density is also obtained with useful precision.
A more detailed discussion of the model dependence of fitted densities
will appear elsewhere.

\acknowledgements
We thank O. Gayou for providing a table of $G_{Ep}$ data and X. Ji and
C. Perdrisat for useful discussions.
The support of the U.S. National Science Foundation under grant PHY-9971819 
is gratefully acknowledged.  

%\bibliographystyle{/group/enp/user/jjkelly/LATEX/prsty}
%\bibliography{/group/enp/user/jjkelly/LATEX/references}

%%%%%%%%% Figures
%\newpage

\newpage
\widetext

\narrowtext
\vfill
\eject

\end{multicols}
\end{document}